\documentstyle[aps,prd]{revtex}

\tighten

\begin{document}

\def \ts {\textstyle}
\def \rd {\displaystyle{\cdot}}
\def \D {\mbox{D}}
\def \p {\partial}
\def \cs {c_{\rm s}^2}

\title{Acoustic oscillations and viscosity}

\author{
Roy Maartens\thanks{email: maartens@sms.port.ac.uk}
}

\address{School of Computer Science and Mathematics, Portsmouth
University, Portsmouth PO1~2EG, Britain}
        
\author{
Josep Triginer\thanks{email: pep@ulises.uab.es}
}

\address{International University of Catalonia, C/Immaculada 22,
08017~Barcelona, Spain} 

\address{and Departament de F\'{\i}sica, Universitat Aut\`onoma de
Barcelona, 08193~Bellaterra, Spain}

\date{July 1998}

\maketitle

\begin{abstract}

Using a simple thermo-hydrodynamic model that respects 
relativistic causality, we revisit the
analysis of qualitative
features of acoustic oscillations in the photon-baryon fluid. 
The growing photon mean free path introduces transient effects
that can be modelled by the causal generalization of relativistic
Navier-Stokes-Fourier theory. Causal thermodynamics
provides a more satisfactory hydrodynamic approximation to 
kinetic theory than the
quasi-stationary (and non-causal) approximations arising from
standard thermodynamics or from
expanding the photon distribution
to first order in the Thomson scattering time.
The causal approach introduces small corrections to the
dispersion relation obtained in quasi-stationary treatments.
A dissipative contribution to the
speed of sound slightly increases the frequency of
the oscillations. The diffusion damping scale
is slightly increased by the causal 
corrections. Thus quasi-stationary approximations tend to
over-estimate the spacing and under-estimate the damping
of acoustic peaks.
In our simple model, the fractional corrections
at decoupling are~$\gtrsim 10^{-3}$.

\end{abstract}


\section{Introduction}

Acoustic oscillations in the photon-baryon fluid before decoupling
leave a vital imprint on the cosmic microwave background \cite{hs}
(and possibly also on large-scale structure \cite{ehss}). 
The form of this imprint encodes information on fundamental
cosmological parameters, and increasingly accurate and refined 
numerical integrations are being performed to produce predictions
that can be tested against current and upcoming observations.
As a complement to detailed numerical models, it is also useful to
analyze qualitatively and analytically the key physical features
such as acoustic oscillations. As pointed out by Hu and 
Sugiyama \cite{hs}, numerical integrations are sufficient for
direct comparison of specific models with observations, but do not
readily produce a qualitative and analytic understanding of the
physical processes at play. In \cite{hs}, acoustic oscillations
and their damping due to photon diffusion are analyzed analytically
via expanding the integrated Boltzmann multipoles 
in the Thomson scattering time $\tau_{_T}$. To zero
order in $\tau_{_T}$, 
i.e. in the tight-coupling approximation, which is valid on
scales much larger than the photon mean free path, the oscillations
are undamped. Damping arises from the first order approximation,
which introduces a shear viscosity via the radiation quadrupole.
This approximation is hydrodynamic, in the sense that all
multipoles beyond the quadrupole are neglected, and it is
effectively equivalent (see \cite{ap}) to the non-equilibrium
thermodynamic approach of Weinberg \cite{w}, which is based on the
relativistic Navier-Stokes-Fourier theory developed by Eckart.

As pointed out by Peebles and Yu \cite{py}, expansions in $\tau_{_T}$,
even beyond first order (see \cite{ap}),
cannot take proper account of the change in the photon mean free time,
especially near to decoupling. In other words, the $O(\tau_{_T})$
approximation, and the equivalent Navier-Stokes-Fourier approximation,
are inherently quasi-stationary, and provide a limited approximation
to the Boltzmann equation. An improved hydrodynamic approximation
to kinetic theory is the causal thermodynamics developed by Israel
and Stewart \cite{is}, which generalizes the non-causal Eckart theory
by incorporating transient non-quasi-stationary effects.
In this paper, we use causal thermodynamics to refine aspects
of the work by Weinberg \cite{w} and
Hu and Sugiyama \cite{hs}. Of course, all hydrodynamic
approximations, which assume that the photon-baryon fluid is
close to equilibrium via interactions, will break down when
the Thomson interaction rate becomes too low, i.e. as decoupling is
approached.

In a previous paper \cite{mt}, we developed a general formalism for
incorporating causal thermodynamics into perturbation theory,
thus providing a self-consistent approach to density
perturbations in dissipative cosmological
fluids. Our formalism is based on the
covariant perturbation theory of Ellis and Bruni \cite{eb,bde}, and
the covariant causal thermodynamics of Israel and Stewart \cite{is}.
Here we apply the general formalism to the case of viscous
damping of density perturbations in the photon-baryon fluid. 
This effect has been comprehensively
analyzed via detailed study of the Boltzmann equation (see e.g.
\cite{py,ks,yss,eh}). Dissipative hydrodynamics provides a
simplified alternative to a full kinetic theory analysis, that can
illuminate some of the key physical effects without the
same level of detail and complexity. 
The approaches of Weinberg \cite{w}
and Hu and Sugiyama \cite{hs}
neglect gravitational effects, which is not unreasonable
on subhorizon scales. We refine their approach by including
metric perturbations. More importantly, their approach is inherently
quasi-stationary, and
the main refinement we introduce is to incorporate
transient effects via the causal generalization
of Navier-Stokes-Fourier thermodynamics.

It turns out that the relaxational effects which are incorporated
into the causal transport equation, but which are neglected in the
non-causal theory, have a small but interesting impact.
The sound speed, which determines the frequency of acoustic 
oscillations for each mode, acquires a positive dissipative
correction. This produces a small increase in the frequency of
acoustic oscillations on each scale. The viscous damping rate
found in quasi-stationary approximations \cite{hs,w}
also acquires a positive correction, leading to a small
increase in the diffusion cut-off scale. A rough estimate based on our
simplified model is that the fractional
corrections are~$\gtrsim 10^{-3}$ by decoupling.

For simplicity, since we are interested in qualitative features
rather than specific numerical predictions, and since we wish
to focus on relaxational corrections to quasi-stationary models,
we assume a flat background universe with
negligible non-baryonic content. 
We treat the mixture of baryonic matter and 
radiation as a single dissipative fluid, with dissipation
arising from the growing mean free path of photons
in their Thomson scattering off electrons. 
We consider the era between
matter-radiation equality and decoupling,
neglecting the matter pressure.
Dissipative effects are expected to be greatest near to
decoupling. 
We will also neglect thermal conduction for simplicity,
while the bulk viscous
stress is necessarily negligible since the fluid is 
in the non-relativistic regime.
Thus the dissipative effect of photon diffusion
is reduced 
(within a radiative hydrodynamic model) 
to a shear viscous stress $\pi_{ab}$.

In non-causal thermodynamics (and in an $O(\tau_{_T})$
expansion of the the Boltzmann multipoles),
this shear viscous stress is
algebraically determined by the shear $\sigma_{ab}$ via \cite{w}
\begin{equation}
\pi_{ab}=-2\eta\sigma_{ab} \,, 
\label{0}\end{equation}
where $\eta$ is the viscosity.
The implicit instantaneous relation between the cause
(shear, i.e. anisotropic expansion rate) and the
effect (anisotropic
stress), is what underlies the pathologies of 
the theory, i.e. that it admits dissipative signals with infinite
wavefront speeds and that its equilibrium states are generically
unstable \cite{hl}.
In the causal theory, the stress is no longer
algebraically and `instantaneously' determined by the shear,
but satisfies an evolution equation in which there is a 
time-lag between cause and effect \cite{is}:
\begin{equation} 
\tau\dot{\pi}_{ab}+\pi_{ab}=-2\eta\sigma_{ab}\,,
\label{1}\end{equation}
where $\tau$ is the relaxation time-scale.
This causal transport
leads to a subluminal speed of viscous pulses, and
introduces small but interesting corrections to the 
quasi-stationary dispersion relation.

Causal thermodynamics has a solid kinetic theory motivation via the
relativistic Grad 14-moment approximation to the Boltzmann
equation \cite{is}. This approximation effectively
restores second-order non-equilibrium terms in the
entropy that are neglected
in the relativistic Chapman-Enskog approximation \cite{d}, which leads
to the non-causal theory. 
Thus causal thermodynamics provides a more satisfactory and less 
incomplete hydrodynamic approximation to kinetic theory than the
standard non-causal theory. 
The differences are greatest for high frequency phenomena, on
scales comparable to the mean free time or path, 
when the Chapman-Enskog
approximation breaks down, while the Grad approximation remains
valid.

We follow the notation of \cite{mt}. Units are such that 
$c=1=8\pi G$, except where numerical values are given in 
specified units.
The present Hubble rate is $H_0=100h$ km/s/Mpc.

\section{Causal viscous perturbations}

Ref. \cite{mt} contains a general
discussion and general covariant perturbation equations for 
causal dissipative
hydrodynamics in cosmology.
In the interests of brevity, we will not repeat here the
derivations and motivations of \cite{mt}, but simply quote
the results as they are needed. 

In the
era between matter-radiation equality and decoupling, the mixture of
photons, baryons and electrons
is treated as a single dissipative fluid,
with dissipation due to photon Silk diffusion \cite{si}.
The baryons and electrons are tightly coupled, culminating
in their recombination, while the
photon coupling to baryonic matter through
Thomson scattering weakens as the mean free path grows.
We assume for simplicity that there is no non-baryonic cold dark
matter. We also assume 
negligible thermal conduction and particle flux , so that 
the particle and energy frames coincide \cite{is}, and we can
choose the fluid four-velocity
$u^a$ (where $u^au_a=-1$) as the four-velocity of this frame.
The bulk viscous stress is negligible, since the fluid is
non-relativistic.

The fundamental quantity in the covariant approach to
density perturbations \cite{eb} is
\begin{equation}
\delta=a\D^a\delta_a=a^2{\D^2\rho\over\rho}\,,
\label{2}\end{equation}
where $\D_a$ is
the covariant spatial derivative, $a$ is the background scale
factor, and
\[
\delta_a=a{\D_a\rho\over\rho}
\]
is the comoving fractional energy density
gradient,
which is a covariant measure of density inhomogeneity.
The total energy density $\rho=\rho_{_B}+\rho_{_R}$
is made up of baryonic and radiation parts, and $\delta$
is the density fluctuation scalar for the photon-baryon
system, considered as a combined single fluid.

Covariant entropy perturbations in a
single dissipative
fluid model (as opposed to a 2-component model),
are sourced by heat flux
and bulk viscous stress, and shear viscous stress has no effect to
linear order
(Eq. (18), \cite{mt}). 
If, as we assume, the initial entropy perturbation is zero,
there are therefore no
entropy perturbations.\footnote{In
any case, the entropy perturbations are
decoupled from the density perturbations, because the
non-barotropic index $(\partial p/\partial s)_\rho$ vanishes
in the background, so that the entropy source term in
the density perturbation equation is zero (Eq. (28), \cite{mt}).}
Then $\delta$
satisfies the evolution equation (Eq. (28), \cite{mt})
\begin{equation}
\ddot{\delta}+2H\left(1-3w+{\ts{3\over2}}
\cs\right)\dot{\delta}-{\ts{3\over2}}H^2\left(1+
8w-3w^2-6\cs\right)\delta-\cs\D^2\delta={\sf S}[\pi]\,,
\label{3}\end{equation}
where an overdot denotes the covariant derivative along $u^a$,
$H=\dot{a}/a$, $w=p/\rho$, with $p=p_{_B}+p_{_R}$ the total
isotropic pressure, and
$\cs=\dot{p}/\dot{\rho}$ is the adiabatic sound speed. The
shear viscous source term is given 
by (Eq. (32), \cite{mt})
\begin{equation}
{\sf S}[\pi]=3H\dot{\cal S}-3H^2(1+6w-3\cs){\cal S}+\D^2{\cal S}\,,
\label{4}\end{equation}
where the covariant shear viscous stress scalar is
\[
{\cal S}=a^2{\D^a\D^b\pi_{ab}\over\rho} \,.
\]
The viscous transport equation (\ref{1}) leads to the following
evolution
equation for ${\cal S}$ (Eq. (46), \cite{mt}):\footnote{The
$\rho$ factors are mistakenly omitted in equations (46) and
(54) of \cite{mt}.}
\begin{equation}
\tau\dot{\cal S}+\left[1-H\left\{3(1+w)\tau-{4\eta\over\rho(1+w)}
\right\}\right]{\cal S}={4\eta\over 3\rho(1+w)}\left[\dot{\delta}
-3wH\delta\right] \,.
\label{6}\end{equation}
Equations (\ref{3})--(\ref{6}) form a coupled system that
governs the evolution of density perturbations with
causal viscosity.

The thermal energy of matter
is much less than the rest mass energy after the epoch $t_{\rm e}$
of matter-radiation equality, so that
$p_{_B}\approx 0$, and then
$p\approx {1\over3}\rho_{_R}$.
Taking $a=1$ at the present epoch, we have in
the background (where matter and radiation are non-interacting)
\begin{equation}
w={1\over3}\left[\left({a\over a_{\rm e}}\right)+1\right]^{-1}<
{1\over6}\,,~~
\cs ={4\over 3}\left[3\left({a\over a_{\rm e}}\right)+4\right]^{-1}
<{4\over 21}\,,
\label{wcs}\end{equation}
for $a>a_{\rm e}$.
To a reasonable approximation (sufficient for the purposes of our 
simple model), we can neglect these quantities relative to 1 in
the round brackets in
equations (\ref{3})--(\ref{6}).
Furthermore, since $w\ll 1$ and $\eta H/\rho\ll 1$,
we can neglect the last term
on the right of Eq. (\ref{6}).

We decompose into Fourier modes $\delta(t,\vec{x})\rightarrow
\Delta(t,\vec{k})$ and ${\cal S}(t,\vec{x})\rightarrow
\Sigma(t,\vec{k})$, where $k$ is the comoving wave number,
so that the proper wave number is $K=k/a$. The proper wavelength
$\lambda=2\pi/K$ is constrained by
$\lambda_{_T}<\lambda<\lambda_{_H}$, 
where $\lambda_{_H}=H^{-1}$ is the Hubble scale (above which 
thermo-hydrodynamic effects do not operate), and
the
minimum scale $\lambda_{_T}$ is the photon mean free
path for Thomson scattering:
\begin{equation}
\lambda_{_T}={1\over n\sigma_{_T}}\,.
\label{5a}\end{equation}
Here
$\sigma_{_T}$ is the Thomson
cross section and $n$ is the free
electron number density. Perturbations on scales 
below $\lambda_{_T}$ will be wiped out by photon streaming
(and the hydrodynamic model breaks down on these scales).
Causal thermodynamics applies down to the Thomson scale,
but the non-causal theory requires $\lambda\gg\lambda_{_T}$.
The damping effect of photon streaming operates beyond the
Thomson mean free scale, since photon diffusion out of
over-dense regions rapidly destroys perturbations below
a critical scale, 
as shown by Silk \cite{si}. The critical
cut-off scale is
the diffusion scale $\lambda_{_D}$, which
is considerably greater than the Thomson mean free
scale, as we will confirm below. 

Collecting the above points, and 
using the background
Friedmann equation $\rho=3H^2$,
the coupled system becomes
\begin{eqnarray}
\ddot{\Delta}+2H\dot{\Delta}-{\ts{3\over2}}H^2\left(1-
{\ts{2\over3}}\cs K_*^2\right)\Delta &=&
3H\dot{\Sigma}-3H^2\left(1+{\ts{1\over3}}
K_*^2\right)\Sigma \,, \label{7}\\
3H\tau\dot{\Sigma}+3H\left(1+\eta_*-3H\tau\right)\Sigma
 &=& \eta_*\dot{\Delta} \,, \label{8}
\end{eqnarray}
where we have defined the dimensionless
expansion-normalized shear viscosity
\begin{equation}
\eta_*={4\eta\over 3H} \,,
\label{9}\end{equation}
and proper wave number
\begin{equation}
K_*={K\over H}=2\pi{\lambda_{_H}\over \lambda}\,.
\label{10}\end{equation}
The relevant scales constrain $K_*$ by
\begin{equation}
2\pi< K_* <  2\pi{\lambda_{_H}\over
\lambda_{_T}}\,.
\label{10b}\end{equation}

Radiative shear viscosity due to Thomson scattering is given
by (see \cite{w}, and \cite{s,a,ui} for refinements
using kinetic theory\footnote{
Note that \cite{s} claims a correction
factor of ${10\over 9}$ in Weinberg's expression for $\eta$.
This arises from anisotropic scattering effects, and a further
small correction is also induced by polarization effects \cite{hs}.
These corrections are not important for our simple model.
})
\begin{equation}
\eta={\ts{4\over15}}r_0T^4\tau_{_T}\,,
\label{12}\end{equation}
where $r_0$ is the
blackbody radiation constant, $T$ is the photon temperature
and $\tau_{_T}=\lambda_{_T}$ is the photon mean free time for
Thomson scattering. 
The causal relaxation time-scale $\tau$ is by Eq. (\ref{1})
the characteristic time taken for the
fluid to relax toward equilibrium if the 
viscous `driving force' were to be
`switched off'. Thus we expect $\tau$ is of the order of a few
Thomson interaction times, and below the characteristic
diffusion time:
\begin{equation}
\tau_{_T}\lesssim \tau<\tau_{_D} \,,
\label{12b}\end{equation}
where $\tau_{_D}=\lambda_{_D}$.

In the era we are considering, $H$ decays approximately 
like $a^{-3/2}$, $n$ decays faster than $a^{-3}$ because of 
recombination effects, 
and $T$ decays as $a^{-1}$. Thus to lowest order
\begin{eqnarray}
K_* &=& (K_*)_{\rm e}\left({a\over a_{\rm e}}\right)^{1/2} \,,
\label{d1}\\
\lambda_{_H}&=& (\lambda_{_H})_{\rm e}\left({a\over
a_{\rm e}}\right)^{3/2} \,, \label{d2}\\
\lambda_{_T}&=& (\lambda_{_T})_{\rm e}\left({a\over
a_{\rm e}}\right)^{3+I} \,, \label{d3}\\
\tau_{_T}H &=& \left({\lambda_{_T}\over
\lambda_{_H}}\right)_{\!{\rm e}}\left({a\over
a_{\rm e}}\right)^{3/2+I} \,, \label{d4}\\
\eta_* &=& (\eta_*)_{\rm e}\left({a\over a_{\rm e}}\right)^{1/2} \,,
\label{d5}
\end{eqnarray}
where $I>0$ encodes the growing effects of recombination on the number
density of free electrons, the details of which are not important
for our simple model.
Using the standard numerical values
\cite{p} of 
$t_{\rm e}\approx 3\times 10^{10} h^{-4}$ sec,
redshift $z_{\rm e}\approx 2\times 10^4h^2$, and 
present baryon number density $n_{_B}\approx
10^8h^8$ cm$^{-3}$, with the present baryonic
(dark and luminescent) density fraction
$\Omega_{_B}\approx 1$,
we find that
\begin{equation}
(\lambda_{_H})_{\rm e}\approx 1.5\times10^{21}h^{-4}\mbox{ cm}  
\,,~(\lambda_{_T})_{\rm e}\approx 10^{16} h^{-8} \mbox{ cm}\,,~
(\eta_*)_{\rm e}\approx 1. 5\times 10^{-6}h^{-4}\,.
\label{5b}\end{equation}

Equations (\ref{7}) and (\ref{8}) govern the evolution of 
density perturbation modes within a 
simplified causal viscous fluid model.
The system may be decoupled to
produce a third-order equation in $\Delta$. 
In practice we analyse the coupled
system, but for completeness, the decoupled equation is
\begin{eqnarray}
&& \tau\left.\ddot{\Delta}\right.^{\!\!\rd}
+\left[1+{\ts{1\over2}}\tau H+\dot{\tau}+F
\right]\ddot{\Delta} \nonumber \\
&&{}~+H\left[2+\left({\ts{3\over2}}+{\ts{1\over3}}
K_*^2\right)\eta_*-{\ts{15\over2}}\tau H+2\dot{\tau}
-(\tau H)^{-1}\left\{\tau\dot{\eta}_*
+F(\eta_*-2\tau H)\right\}\right]\dot{\Delta} \nonumber \\
&&{}~-{\ts{3\over2}}H^2\left[\left(1+\eta_* -
{\ts{7\over2}}\tau H+\dot{\tau}+F\right)
\left(1-{\ts{2\over3}}\cs K_*^2\right)-\tau H-{\ts{2\over3}}\tau K_*^2
\left(\cs\right)^{\rd}\right]\Delta =0 \,,
\label{10a}
\end{eqnarray}
where
\[
F\equiv -\tau\left[\ln \left|\tau H\left(6-K_*^2\right)
-3(1+\eta_*)\right|\right]^{\rd} \,,
\]
and we used the background field equation $\dot{H}=-{3\over2}H^2$.
In the non-causal case $\tau=0$, Eq. (\ref{10a}) reduces to
the second order equation
\begin{equation}
 \ddot{\Delta}+H\left[2+\left({\ts{3\over2}}+{\ts{1\over3}}K_*^2
\right)\eta_*-{\dot{\eta}_*\over H(1+\eta_*)}\right] \dot{\Delta}
-{\ts{3\over2}}H^2\left[
(1+\eta_*)\left(1-{\ts{2\over3}}\cs K_*^2\right)\right]
\Delta =0 \,.
\label{10h}\end{equation}

\section{Viscous acoustic oscillations}

The key features of the solutions are readily analyzed qualitatively,
given the simplicity of our model. 
We assume that the coefficients in equations (\ref{7}) and (\ref{8})
are slowly varying, which is reasonable for small-scale modes.
Then we can try a WKB solution of form
$\Delta(t,k)=a^{-1}A(k)\exp i\int\omega\,dt$ and 
$\Sigma(t,k)=a^{-1}B(k)\exp i\int\omega\,dt$, where the scale
factor terms remove
the non-dissipative effects of expansion on small-scale
modes, facilitating comparison with Weinberg's results, which neglect
expansion.
This gives
\[
\left[
\begin{array}{lll}
-\omega^2-{\ts{1\over2}}H^2\left(5- 
2\cs K_*^2\right) &~~& -3iH\omega+(6+K_*^2)H^2 \\
{}&{}&\\
-i\eta_*\omega+\eta_*H &~~& 3i\tau H\omega+3(1+\eta_*-4\tau H)H \\
\end{array}
\right]\left[
\begin{array}{c}
A\\
{}\\
B\\
\end{array}
\right]=0 \,.
\]
Non-trivial solutions require a vanishing determinant of the
coefficient matrix, leading to the dispersion equation
\begin{eqnarray}
&& 2i\tau \omega^3+2(1-4\tau H)\omega^2 
-i\left[2\eta_*\left(2
+{\ts{1\over3}}K_*^2\right)+\tau H\left(2\cs K_*^2-5\right)
\right]H\omega \nonumber\\
&&{}~
+\left[ \left(1+\eta_*-4\tau H\right)\left(5-2
\cs K_*^2\right)+2\eta_*\left(1+{\ts{1\over3}}K_*^2\right)\right] 
H^2=0 \,.
\label{10c}\end{eqnarray}
In general, the solutions will have the form
\begin{equation}
\omega= \pm Kv+i\,\Gamma \,,
\label{10e}\end{equation}
where $v$ is the (real) phase speed 
of oscillations and $\Gamma$ is the viscous damping rate.
Damping will be severe for scales $\lambda<\lambda_{_D}$,
where the diffusion cut-off scale is defined by equality of the
damping rate and the expansion rate, i.e.
$\Gamma(\lambda_{_D})=H$.

In the dispersion relation (\ref{10c}), we can neglect $\eta_*$
relative to 1 by equations (\ref{d5}) and (\ref{5b}).
Equations (\ref{wcs}), (\ref{10}), (\ref{d2}) and (\ref{5b}) show
that for the relevant
small scales ($\lambda\ll \lambda_{_H}$), we have\footnote{
Numerical integrations in the special
case $\cs K_*^2\ll 1$ are given in \cite{t}.
}
$\cs K_*^2\gg 1$.
Then the dispersion cubic may be written
\begin{equation}
i\tau_* \omega_*^3+(1-4\tau_*)\omega_*^2 
-i\left(
{\ts{1\over3}}\eta_*K_*^2+\tau_*\cs K_*^2
\right)\omega_* - \left(1-4\tau_*\right)
\cs K_*^2 =0 \,,
\label{dis}\end{equation}
where $\omega_*=\omega/H$ and $\tau_*=\tau/H^{-1}$ are dimensionless
expansion-normalized variables.
For the non-causal theory, with $\tau=0$, Eq. (\ref{dis})
becomes
\begin{equation}
3\bar{\omega}_*^2-\left(i\,\eta_*K_*^2\right)
\bar{\omega}_*-3\cs K_*^2=0 \,,
\label{16}\end{equation}
where an overbar indicates a
quantity in the non-causal theory. Eq. (\ref{16}) has
solution
\[
\bar{v}^2=\cs\left[1-{\bar{\Gamma}^2\over \cs K^2}\right]\,,~~
\bar{\Gamma}={\ts{1\over6}}\eta_*K_*^2H\,.
\]
Since $\bar{\Gamma}/H<1$ for the scales that survive viscous damping,
and $\cs K_*^2\gg 1$ for the relevant sub-horizon scales,
we have $\bar{v}^2=\cs$ to a good approximation. Thus
\begin{equation}
\bar{v}=c_{\rm s}\,,~~~\bar{\Gamma}
=\left({2\eta\over3\rho}\right)K^2\,,
\label{14}\end{equation}
which agrees (when the isotropic pressure, heat flux and bulk stress
are negligible)
with Weinberg's special relativistic generalization
of the non-relativistic
Landau-Lifshitz result \cite{w}.\footnote{See also
\cite{g} for a kinetic theory approach, and
\cite{sb} for the inclusion of magnetic
fields.}
The diffusion cut-off scale follows from Eq. (\ref{14}) 
as
\begin{equation}
\bar{\lambda}_{_D}={2\pi\over3}\sqrt{2\eta\over H}\,
\lambda_{_H}\,.
\label{14a}\end{equation}
By equations (\ref{d2}), (\ref{d3}) and (\ref{d5}),
the diffusion scale 
is well above the Thomson scale, although
the ratio decreases with expansion: 
\begin{equation}
{\bar{\lambda}_{_D}
\over\lambda_{_T}}\approx 500h^2 \left({a_{\rm e}
\over a}\right)^{5/4+I}\,.
\label{15}\end{equation}

Now we consider the causal corrections to Eq. (\ref{14}). From
equations (\ref{12b}), (\ref{d4}), (\ref{5b}) and (\ref{14a}),
it follows that $\tau_*\equiv\tau H\ll 1$.
We write
$\omega_*=\bar{\omega}_*(1+y\tau_*)$, where $\bar{\omega}_*$ satisfies
Eq. (\ref{16}), and solve the dispersion cubic
Eq. (\ref{dis}) to first order in $\tau_*$. This gives
\[
y={\bar{\Gamma}\over H}\left[1+i\,{(\bar{\Gamma}+4H)
\over \pm c_{\rm s}K}\right]\,,
\]
on using $\cs K_*^2\gg 1$. It follows that
\begin{eqnarray}
v &=& c_{\rm s}\left(1+\tau\bar{\Gamma}\right) \,, \label{18a}\\
\Gamma &=& \bar{\Gamma}\left[1+2\tau\left(\bar{\Gamma}+2H
\right)\right] \,. \label{18b}
\end{eqnarray}
Thus the causal generalization of standard thermodynamics
leads to corrections that increase the effective sound speed
and damping rate.
The latter increase leads to an increase in the diffusion cut-off
scale, which is determined by setting $\bar{\Gamma}\approx H$
in Eq. (\ref{18b}):
\begin{equation}
\lambda_{_D}=\bar{\lambda}_{_D}\left(1+3\tau H\right) \,.
\label{19}\end{equation}

In order to estimate the size of the corrections, we need 
an expression for the causal relaxation timescale $\tau$. 
In the case of a simple (i.e. single-species)
relativistic gas (obeying classical or quantum 
statistics), the thermodynamic coefficients such as
$\tau$ and $\eta$ are in principle known via complicated
kinetic theory
formulas involving integrals over 
energy and collision cross-sections \cite{is}. 
In the radiative transfer context, where 2 particle species are
involved, one treated hydrodynamically, 
the shear viscosity is given by Eq. (\ref{12}), while $\tau$ is
implicitly assumed to be zero in quasi-stationary 
treatments \cite{w,s,a}. If Grad-type moment approximations
are used, then
$\tau$ is equal to the mean collision time as a consequence
of the hydrodynamic assumptions \cite{ui,sch}.
However, it seems physically reasonable that
$\tau$ could be greater than $\tau_{_T}$, if the hydrodynamic
assumptions in radiative transfer are somewhat relaxed.
Thus a simple generalization
\begin{equation}
\tau= \alpha \tau_{_T} \,,
\label{20}\end{equation}
where $\alpha\gtrsim 1$ is a constant, seems reasonable within
the phenomenology of thermodynamics.

The correction to the diffusion scale, and the maximal correction
to the sound speed (which occurs at $\bar\Gamma=H$), are determined
by $\tau H$. By equations (\ref{20}), (\ref{d4}) and (\ref{5b}),
we find that in our simple model
\begin{equation}
\tau H=\alpha \left({\lambda_{_T}\over \lambda_{_H}}\right)_{\rm e}
\left({a\over a_{\rm e}}\right)^{3/2+I}\approx 
 7 \alpha\times 10^{-6}h^{-4}\left({a\over a_{\rm e}}
\right)^{3/2+I}\,.
\label{21}\end{equation}
The correction factor grows with expansion, as expected, since the
mean free photon path is growing. We take $h=0. 6$.
At matter-radiation equality, the fractional
correction is then 
$\approx 5\alpha\times 10^{-5}\gtrsim 5\times 10^{-5}$. 
As decoupling is approached, 
the ionization parameter $I$ becomes more important. However, for
a rough under-estimate, we can set $I=0$.
(In any case, all hydrodynamically
based models will break down near decoupling.) With $z_{\rm dec}
\approx 1100$, we find that the maximal fractional correction
introduced by relaxational effects is
\begin{equation}
\left(\tau H\right)_{\rm max} > \alpha\times 10^{-3}
\gtrsim 10^{-3}\,,
\label{22}\end{equation}
in our simple model. This is the relevant estimate, since the imprint
of acoustic oscillations is frozen in at decoupling.

\section{Conclusions}

Our principal result is the causal generalization and upward
correction of the quasi-stationary
sound speed, given in Eq. (\ref{18a}),
and damping rate, given in Eq. (\ref{18b}).
These corrections are small, at roughly the $0.1$--$1$~\%
level, as given in Eq. (\ref{22}), but they 
show how quasi-stationary approximations tend to under-estimate
the frequency and damping of acoustic oscillations, and they
indicate
the refinements that would arise from a more complete kinetic theory
approach, which involves far greater computational complexity.
The qualitative features of the
refinements are: (a) 
a higher threshold for the survival of density
perturbations, i.e. an increased diffusion cut-off scale, given
by Eq. (\ref{19}); (b) the appearance
of a dissipative contribution to the sound speed of the photon-baryon
fluid (consistent with the general theoretical results for
perturbations of stationary fluids in flat spacetime \cite{is,hl})
which increases the frequency of oscillations;
(c) the scale-dependence of the corrections to the damping rate
and sound speed;
(d) the central role played by the
relaxation time-scale $\tau$, which provides a simple
scalar encoding for the non-quasi-stationary
effects of photon diffusion.

This paper shows that in principle
dissipative hydrodynamics with causal
transport equations can provide an economical alternative to 
kinetic theory models, with some additional insights into the
nature and source of acoustic oscillations.
The simplified model used here
could be refined, principally by incorporating 
thermal conduction, to produce an improved model that
nevertheless will still allow a qualitative and analytic treatment,
unlike the full kinetic theory approach.

Our model can also be seen as part of
a growing body of examples in 
cosmology (e.g. \cite{bk}),
relativistic astrophysics (e.g. \cite{dh}), and
other areas of physics (e.g. \cite{jcl}),\footnote{In
non-relativistic physics, causal theories are usually called
`extended' or `hyperbolic'.}
which show that interesting and sometimes
significant physical differences can
arise from the causal approach to thermodynamics. 
(A general overview is
given in \cite{apr}.)
Even if the causal corrections are small, as is the case here,
they provide further insight into the physics, with the added
advantage that
one avoids the essentially unsatisfactory
features of non-causal thermodynamics. The predictions of
the non-causal theory are
readily recovered in the appropriate limit.

\acknowledgements

JT was supported by the DGICYT under Grant
No. PB94-0718, and would also like to thank
the Relativity and 
Cosmology Group at Portsmouth University, 
where part of this work was done, for hospitality.
\\

\end{document}